\documentclass[12pt]{article}
\usepackage{amsmath}
\usepackage{amssymb}
\usepackage{latexsym}
\usepackage{hyperref}

\def\braket#1{\mathinner{\langle{#1}\rangle}}

\def\ket#1{\left|#1\right>}
{\catcode`\|=\active 
  \gdef\Braket#1{\left<\mathcode`\|"8000\let|\bravert {#1}\right>}}
\def\bravert{\egroup\,\vrule\,\bgroup}
\newcommand{\alg}[1]{\mathfrak{#1}}

\newcommand{\su}{\alg{su}}
\newcommand{\psu}{\alg{psu}}
\newcommand{\Sl}{\alg{sl}}
\newcommand{\so}{\alg{so}}

\newcommand{\tr}{\mathop{\rm tr}}

\newcommand{\be}{\begin{eqnarray}}
\newcommand{\ee}{\end{eqnarray}}
\newcommand{\bea}{\begin{eqnarray}}
\newcommand{\eea}{\end{eqnarray}}
\newcommand{\ben}{\begin{equation}}
\newcommand{\een}{\end{equation}}

\newcommand{\del}{\partial}
\newcommand{\Qo}{\widehat Q}

\newcommand{\nn}{\nonumber}

\numberwithin{equation}{section}

\begin{document}

\begin{titlepage}
\begin{flushright}
CALT-68-2527\\
hep-th/0410282
\end{flushright}
\vspace{15 mm}
\begin{center}
{\huge	Quantum string integrability 
	and AdS/CFT }
\end{center}
\vspace{12 mm}

\begin{center}
{\large Ian Swanson}\\
\vspace{3mm}
California Institute of Technology\\
Pasadena, CA 91125, USA 
\end{center}
\vspace{5 mm}
\begin{center}
{\large Abstract}
\end{center}
\noindent

Recent explorations of the AdS/CFT correspondence have unveiled integrable structures 
underlying both planar ${\cal N}=4$ super-Yang-Mills theory and type IIB string theory
on $AdS_5\times S^5$.  Integrability in the gauge 
theory emerges from the fact that the dilatation generator can be identified with 
the Hamiltonian of an integrable quantum spin chain, and the classical
string theory has been shown to contain infinite towers of hidden
currents, a typical signature of integrability.  
Efforts to match the integrable structures of various classical string 
configurations to those of corresponding gauge 
theory quantum spin chains have been largely successful. 
By studying a semiclassical expansion about a class of point-like solitonic solutions to the 
classical string equations of motion on $AdS_5\times S^5$,
we take a step toward demonstrating that integrability in the string theory survives quantum 
corrections beyond tree level.  Quantum fluctuations are chosen to align with 
background curvature corrections to the pp-wave limit of $AdS_5\times S^5$, 
and we present evidence for an infinite tower of local bosonic charges that are conserved
by the quantum theory to quartic order in the expansion.  
We explicitly compute several higher charges based on a Lax representation of the 
worldsheet sigma model and provide a prescription for matching the eigenvalue spectra of these
charges with corresponding quantities descending from the integrable structure of the 
gauge theory. 

\vspace{1cm}
\begin{flushleft}
\today
\end{flushleft}
\end{titlepage}
\newpage
\section{Introduction}
The emergence of integrable structures from planar ${\cal N}=4$ super-Yang-Mills (SYM)
theory and type IIB string theory on $AdS_5\times S^5$ has renewed hope that 
't~Hooft's formulation of large-$N_c$ QCD may eventually lead to an exact solution.
If both the gauge and string theories are in fact integrable,
each will admit infinite towers of hidden charges and,
analogous to the usual identification of the string theory Hamiltonian with 
the gauge theory dilatation generator, there will be an infinite number of mappings 
between the higher hidden charges of both theories.  
This has led to many novel tests of the AdS/CFT correspondence, particularly in the context of
the pp-wave/BMN limits \cite{Berenstein:2002jq,Metsaev:2001bj,Metsaev:2002re}. 
Barring an explicit solution, one would hope that both theories will at least be shown to 
admit identical Bethe ansatz equations, allowing us to explore a much larger 
region of the gauge/string duality.

The fact that the gauge theory harbors integrable structures 
was realized by Minahan and Zarembo when they discovered that a particular 
$SO(6)$-invariant sector of the SYM dilatation generator can be mapped,
at one-loop order in the 't~Hooft coupling $\lambda = g_{\rm YM}^2 N_c$, to the 
Hamiltonian of an integrable quantum spin chain with $SO(6)$ vector lattice sites 
\cite{Minahan:2002ve}.  
The Hamiltonian of this system can be diagonalized by solving a set of algebraic Bethe ansatz
equations:  
the problem of computing operator anomalous dimensions in this sector of the 
gauge theory was thus reduced in \cite{Minahan:2002ve} to solving the
set of Bethe equations specific to the $\so(6)$ sector of the theory.
The correspondence between operator dimensions and integrable 
spin-chain systems at one loop in $\lambda$ was extended to include the complete 
$\psu(2,2|4)$ superconformal symmetry algebra of planar ${\cal N}=4$ SYM theory by Beisert and 
Staudacher in \cite{Beisert:2003yb}. 
Studies of higher-loop integrability in the gauge theory were advanced 
in \cite{Serban:2004jf,Beisert:2004hm},
where so-called long-range Bethe ansatz equations, which are understood
to encode interactions on the spin lattice that extend beyond nearest-neighbor sites,  
were developed for a closed bosonic $\su(2)$ sector of the 
gauge theory.  In this context, closure refers to the fact that operators in this sector
are guaranteed by group theoretical constraints to not mix with other operators in the theory
under the action of the dilatation generator (to all orders in $\lambda$) 
\cite{Beisert:2003tq,Beisert:2003jj}, 
and such sectors are typically labeled by the subalgebra of the full superconformal 
symmetry algebra under which they are invariant.  
(The spectral predictions provided by the long-range ansatz of \cite{Beisert:2004hm} were checked against
an alternative virial technique in \cite{Virial}, and agreement was obtained to a high 
degree of precision.)  The dynamics of the gauge theory therefore appear to be copasetic with the 
expectations of integrability, at least to three-loop order in the 't~Hooft expansion, and 
there is convincing evidence that this extends to even higher order \cite{Beisert:2004hm,Beisert:2004ry}.

Concurrent with the introduction 
of the Bethe ansatz formalism in the $\so(6)$ sector of the gauge theory \cite{Minahan:2002ve}, 
related developments emerged from studies of semiclassical configurations of rotating 
string on $AdS_5\times S^5$.  This branch of investigation began
with \cite{Gubser:2002tv}, where the pp-wave limit of the string theory
was reinterpreted in the context of a semiclassical expansion about certain solitonic
solutions in the full $AdS_5\times S^5$ target space.  Using this 
semiclassical picture, Frolov and Tseytlin computed a class of two-spin string solutions
in \cite{Frolov:2002av},
demonstrating explicitly how stringy corrections in the large-spin limit
give rise to systems that can be understood as generalizations of the original pp-wave 
solution studied in \cite{Berenstein:2002jq,Metsaev:2001bj,Metsaev:2002re}.  
This work was extended by a more general study of multi-spin string solutions 
in \cite{Frolov:2003qc}, where the authors provided a detailed prescription 
for making direct comparisons with perturbative gauge theory.  
(For a more complete review of the development and current status of semiclassical
string theory and the match-up with gauge theory, see \cite{Tseytlin:2003ii} and references therein.)
Early indications of integrability in the classical limit of the string theory emerged when 
it was shown that a certain configuration of the Green-Schwarz superstring 
action on $AdS_5\times S^5$ admits an infinite set of classically
conserved non-local charges, and may therefore be an integrable theory 
itself \cite{Bena:2003wd} (see also \cite{Alday:2003zb} for a reduction to the pp-wave system). 
The gauge theory analogue of this non-local symmetry 
was studied in \cite{Dolan:2004ps,Dolan:2003uh},
where a direct connection with the string analysis was made to one-loop
order in $\lambda$.  Various subtleties surrounding studies of the non-local (or Yangian) algebra 
arise at higher loops, and further work is certainly warranted.

In addition to the sector of non-local charges, however, integrable systems typically 
admit an infinite tower
of local, mutually commuting charges, each of which are diagonalized by a set of Bethe 
equations \cite{Faddeev:1996iy,Faddeev:ph}.   
The presence of such a sector of hidden, classically conserved bosonic charges in the 
string theory was pointed out in \cite{Mandal:2002fs}.
Moreover, in accordance with the expectations
of AdS/CFT duality, various studies have been successful in matching
hidden local charges in the classical string theory to corresponding quantities
in the quantum spin-chain formulation of ${\cal N}=4$ SYM theory.  
In \cite{Arutyunov:2003rg}, for example,
Arutyunov and Staudacher constructed an infinite series of conserved local charges in the 
bosonic string theory by solving the B\"acklund equations associated with certain extended 
classical solutions of the $O(6)$ string sigma model.  The local charges generated 
by the B\"acklund transformations were then matched to corresponding conserved charges obtained 
from an integrable quantum spin chain on the gauge theory side.  In fact, they were able to 
demonstrate agreement between both sides of the duality for the entire infinite tower of 
local commuting charges.   This study was extended
in \cite{Arutyunov:2003uj,Arutyunov:2003za}, where it was shown that a general 
class of rotating classical string solutions can be mapped to solutions of a 
Neumann (or Neumann-Rosochatius) integrable system.  
More recently, a class of three-spin classical string solutions was shown in 
\cite{Engquist:2004bx} to generate hidden local charges 
(again via B\"acklund transformations) that match their gauge theory counterparts to one-loop order.  
(For a thorough review of the match-up of semiclassical string
integrable structures with corresponding structures in the gauge theory, see also 
\cite{Beisert:2004ry,Tseytlin:2003ii}.)

The mapping between string and gauge theory integrable structures was studied from a somewhat different
perspective in \cite{Kazakov:2004qf}, where it was shown that the generator of local, classically 
conserved currents
in the string theory is related in certain sectors to a particular Riemann-Hilbert problem
which is reproduced precisely by the gauge theory integrable structure at one and two loops in $\lambda$.  
An analogous treatment of the corresponding Riemann-Hilbert problem in 
non-compact sectors of the gauge/string
duality was carried out in \cite{Kazakov:2004nh}, and an extension of these studies to a larger $\so(6)$ 
sector was recently achieved in \cite{Beisert:2004ag}.
The structure of the higher-loop Riemann-Hilbert problem descending from the classical string theory 
and its relationship with the corresponding gauge theory problem was used in conjunction with 
the long-range gauge theory Bethe ansatz of \cite{Beisert:2004hm} 
to develop an ansatz which, albeit 
conjecturally, is purported to interpolate between the classical and quantum regimes 
of the string theory \cite{Arutyunov:2004vx}.  Although this proposal is 
not a proof of quantum integrability on the string side, it was demonstrated in 
\cite{McLoughlin:2004dh} that the quantized string theory
in the near-pp-wave limit yields a general multi-impurity spectrum that matches the
string Bethe ansatz spectrum of \cite{Arutyunov:2004vx}.  
The intricacy of this match-up is quite remarkable, and stands as strong 
evidence that this ansatz is correct for the string theory, at least to $O(1/J)$ in the 
large angular momentum (or background curvature) expansion.  
Furthermore, the proposed string Bethe equations can accommodate the 
strong-coupling $\lambda^{1/4}$ scaling behavior predicted in \cite{Gubser:1998bc}.  The
spin chain theory implied by these Bethe equations, however, appears to disagree with that of 
the gauge theory, even at weak coupling \cite{Beisert:2004jw}.

Although the Bethe equations of \cite{Arutyunov:2004vx} reproduce several predictions of 
the string theory in a highly nontrivial way, a direct test of quantum 
integrability (beyond tree level) in the string theory is still needed:  this is the
intent of the present work.  Early steps in this direction were taken in \cite{Swanson:2004mk},
where the presence of a conserved local charge responsible for a certain parity degeneracy 
in the near-pp-wave string spectrum was examined at sixth-order in field fluctuations, 
or at $O(1/J^2)$ in the large-$J$ expansion.  Various subtleties of the analysis (possibly involving
the proper renormalization of the theory at $O(1/J^2)$ in the expansion) made it difficult to
reach any concrete conclusions, however.  In this paper we take a more immediate approach,
relying primarily on a Lax representation of the classical string sigma model
and studying a semiclassical expansion about certain point-like solitonic solutions.   
The goal is to establish the existence of a series of conserved, mutually commuting
charges in the string theory that can be quantized and studied using first-order
perturbation theory.  By aligning field fluctuations with the 
finite-radius curvature expansion in \cite{Parnachev:2002kk,Callan:2003xr,Callan:2004uv,Callan:2004ev}, 
we are able to study quantum corrections to quartic order, 
or to one loop beyond tree level.  We show directly that several of the low-lying 
hidden charges in the series are conserved by the 
quantum theory to this order in the expansion, and we propose a method for matching
specific eigenvalues of these charges to corresponding spectral quantities in the gauge theory.

The paper is organized as follows.  
In Section 2 we review the procedure for string quantization in 
the near-pp-wave limit developed in \cite{Callan:2003xr,Callan:2004uv}, 
and we demonstrate how background curvature corrections to the pp-wave theory can be
interpreted as quantum corrections in a particular semiclassical expansion about
point-like classical string solutions.  In Section 3 we show how a
Lax representation of the $O(4,2)\times O(6)$ nonlinear sigma model can be 
modified to encode the string dynamics to the order of interest in this semiclassical
expansion.  We then generate a series of hidden local charges by expanding a perturbed
monodromy matrix of the Lax representation in powers of the spectral parameter.
In section 4 we compute the eigenvalues of these charges in certain 
protected subsectors of the theory in the space of two-impurity string states.
The resulting spectra are then compared on the $S^5$ subspace with those of 
corresponding charges descending
from the $\su(2)$ integrable sector of the gauge theory.  We provide a prescription for
matching the spectra of local charges on both sides of the duality, and carry out 
this matching procedure to eighth order in the spectral parameter.  To the extent that they can be compared
reliably, the gauge and string theory predictions are shown to match to this order 
(and presumably continue to agree at higher orders).  We are thus led to believe that the 
integrable structure of the classical string theory survives quantization, 
at least to the first subleading order in field fluctuations beyond tree level.
We conclude in the final section by outlining future directions of study.

\section{Semiclassical string quantization in $AdS_5\times S^5$}
Most of the literature comparing semiclassical bosonic string theory in $AdS_5\times S^5$
to corresponding sectors of gauge theory operators has focused on classical extended string solutions
to the worldsheet sigma model in either ``folded'' or ``circular'' configurations, 
where certain components of the string angular momentum (ie.~certain charges of the Cartan 
subalgebra of the global symmetry group) are taken to be large (see, eg.~\cite{Frolov:2002av,Frolov:2003qc,Arutyunov:2003uj}).
The latter amounts to choosing a so-called ``spinning ansatz'' for the string 
configuration 
\cite{Frolov:2002av,Frolov:2003qc,Tseytlin:2003ii,Mandal:2002fs,Arutyunov:2003uj,Arutyunov:2003za}, 
and solutions endowed with such an ansatz
can be identified with periodic solutions of the Neumann
(or Neumann-Rosochatius) integrable system.
The standard bosonic worldsheet action is usually chosen with flat worldsheet metric
so that it is easily rewritten in terms of ${\bf R}^6$ embedding coordinates 
and identified with an $O(4,2)\times O(6)$ sigma model.  
In the present study we will modify this treatment to allow for curvature
corrections to the worldsheet metric, a complication that we are forced
to confront when moving beyond tree level in lightcone gauge \cite{Callan:2003xr,Callan:2004uv}.

We begin with a particular form of the $AdS_5\times S^5$ target space metric, 
chosen originally in \cite{Callan:2003xr,Callan:2004uv} for the fact that it admits a simple
form for the spin connection:
\begin{equation}
\label{metric}
ds^2_{AdS_5\times S^5}  =  R^2
\biggl[ -\left({1+ \frac{1}{4}z^2\over 1-\frac{1}{4}z^2}\right)^2dt^2
        +\left({1-\frac{1}{4}y^2\over 1+\frac{1}{4}y^2}\right)^2d\phi^2
    + \frac{d z_k dz_k}{(1-\frac{1}{4}z^2)^{2}}
    + \frac{dy_{k'} dy_{k'}}{(1+\frac{1}{4}y^2)^{2}} \biggr]~.
\end{equation}
While we will not address fermions in this study, we will eventually return to the 
crucial issues of supersymmetry, and the metric choice in 
eqn.~(\ref{metric}) will undoubtedly simplify further investigations.
Here, the bosonic coordinates $z_k$ and $y_{k'}$ span a transverse $SO(4)\times SO(4)$
space, with $j,k,l,\ldots \in 1,\ldots,4$  and $j',k',l',\ldots \in 5,\ldots,8$.  
The $z_k$ coordinates are always chosen to lie in the $AdS_5$
subspace, while $y_{k'}$ are coordinates on $S^5$. 
The scale factor $R$ is the common radius of both subspaces.  
By defining
\be
\cosh\rho \equiv \frac{1+\frac{1}{4}z^2}{1-\frac{1}{4}z^2}\qquad 
\cos\theta \equiv \frac{1-\frac{1}{4}y^2}{1+\frac{1}{4}y^2}\ ,
\ee
we may write the ${\bf R}^6\times {\bf R}^6$ embedding coordinates of $AdS_5$ and $S^5$ as
\be
	Z_{k}  =  \sinh\rho\,\frac{z_{k}}{||z||} &\qquad& Z_0 + iZ_5  =  \cosh\rho\, e^{it}\ ,
\nn\\
        Y_{k'}  =  \sin\theta\,\frac{y_{k'}}{||y||} &\qquad& Y_5 +i Y_6  =  \cos\theta\, e^{i\phi}\ ,
\ee
with $||z|| \equiv \sqrt{z_{k} z_{k}}$.  The coordinates $Z_P$, with $P,Q=0,\ldots,5$, parameterize
$AdS_5$ and are contracted over repeated indices using the metric 
$\eta_{PQ}=(-1,1,1,1,1,-1)$.  The coordinates $Y_M$, with $M,N=1,\ldots,6$, encode 
the $S^5$ geometry, and are contracted with a Euclidean metric.

Decomposing the theory into $AdS_5$ and $S^5$ subspaces,
the usual conformal-gauge worldsheet action
\be
S = -\int d^2\sigma\,  h^{ab} G_{\mu\nu} \del_a x^\mu \del_b x^\nu\ 
\label{action1}
\ee
can be written as 
\be
S  &=&  \int d^2\sigma ({\cal L}_{AdS_5} + {\cal L}_{S^5} )\ ,
\nn\\
{\cal L}_{AdS_5} & = &  -\frac{1}{2} h^{ab} \eta_{PQ}\del_a Z_P \del_b Z_Q  
	+ \frac{\tilde\varphi}{2} \left(\eta_{PQ} Z_P Z_Q + 1\right)\ ,
\label{action2a}
\\
{\cal L}_{S^5} & = &  -\frac{1}{2}h^{ab} \del_a Y_M \del_b Y_M  
	+ \frac{\varphi}{2} \left(Y_M Y_M - 1\right) \ .
\label{action2b}
\ee
The quantities $\varphi$ and $\tilde \varphi$ act as 
Lagrange multipliers in the action, enforcing 
the following conditions:\footnote{Note that, in general, $\varphi$ and $\tilde \varphi$
will depend on dynamical variables.  We thank Arkady Tseytlin for clarification on this point.}
\be
\eta_{PQ} Z_P Z_Q = -1\ , \qquad Y_M Y_M = 1\ .
\ee
The action in eqn.~(\ref{action1}) must also be supplemented by the
standard conformal gauge constraints, and the worldsheet metric $h^{ab}$ (the worldsheet indices 
run over $a,b \in \tau,\sigma$) will be allowed to acquire curvature corrections in accordance with these 
constraints.

We wish to study a semiclassical expansion about the following classical point-like 
(or ``BMN-like'') solutions to the sigma model equations of motion:
\be
t = \phi = p_-\tau \qquad z_k = y_{k} = 0\ .
\label{soliton}
\ee
The expansion is defined in terms of quantum field fluctuations 
according to the following rescaling prescription:
\begin{eqnarray}
\label{rescalePre}
    t \rightarrow x^+
\qquad
    \phi \rightarrow x^+ + \frac{x^-}{\sqrt{\xi}}
\qquad
    z_k \rightarrow \frac{z_k}{\xi^{1/4}}
\qquad
    y_{k} \rightarrow \frac{y_{k}}{\xi^{1/4}}\ .
\end{eqnarray}
(A similar but notably different choice was made in \cite{Frolov:2002av}.)
This particular choice of lightcone coordinates will allow us to maintain a
constant momentum distribution on the worldsheet.  
Additionally, as noted in \cite{Callan:2003xr,Callan:2004uv}, 
it will have the effect of eliminating all normal-ordering 
ambiguities from the resulting worldsheet theory, an outcome which is particularly desirable in
the present study.  Furthermore, we note that if we identify $\xi \equiv R^4$, the proposed 
expansion about the classical solution in eqn.~(\ref{soliton}) 
is identical to the large-radius curvature expansion about the pp-wave limit
of $AdS_5\times S^5$ studied in \cite{Callan:2003xr,Callan:2004uv,Callan:2004ev}.  
In other words, we have chosen a 
perturbation to the classical point-like string geodesic that reproduces the 
target-space curvature perturbation to the pp-wave limit.  
The background metric in eqn.~(\ref{metric}) thus yields the following large-$R$ expansion:
\be
ds^2 & = & 2dx^+ dx^- - (x^A)^2 (dx^+)^2 + (dx^A)^2 
\nonumber \\
& & 	+ \frac{1}{R^2}\left[ 
	-2y^2 dx^+ dx^- + \frac{1}{2}(y^4-z^4) (dx^+)^2 + (dx^-)^2
	+\frac{1}{2}z^2 dz^2 - \frac{1}{2}y^2 dy^2 \right] 
\nonumber \\
\label{expmetric2}
& & 	+ {O}\left(R^{-4}\right),
\label{metricexp}
\ee
where the pp-wave geometry emerges at leading order, and 
$x^A$ are transverse $SO(8)$ coordinates, with $A \in 1,\ldots,8$.

The details of quantizing the string Hamiltonian in this setting
are given in \cite{Callan:2003xr,Callan:2004uv}
(see also \cite{Kallosh:1998zx,Kallosh:1998ji,Metsaev:1998it,Metsaev:1999gz,Metsaev:2000yf}
for further details), 
though we will briefly review the salient points here.
The lightcone Hamiltonian $H_{\rm LC}$ is the generator of worldsheet time translations, 
and is defined in terms of the Lagrangian by
\be
-H_{LC} = -p_+ = \delta {\cal L} / \delta \dot x^+\ ,
\ee 
(or $\Delta -J$ in the language of BMN), and   
this variation is performed prior to any gauge fixing.  
The non-physical lightcone variables $x^\pm$ are 
removed from the Hamiltonian by fixing lightcone gauge $x^+ = p_-\tau$ and replacing 
$x^-$ with dynamical variables by enforcing the conformal gauge constraints
\be
T_{ab} = \frac{\delta {\cal L}}{\delta h^{ab}} = 0\ .
\label{CGC}
\ee 
This procedure can be defined order-by-order in the large-$R$ expansion.  
At leading order, for example, we obtain the following from eqn.~(\ref{CGC}):
\be
\dot x^- & = & \frac{p_-}{2}(x^A)^2 - \frac{1}{2p_-}\left[(\dot x^A)^2 + ({x'}^A)^2\right]
	+ O(1/R^2)\ ,
\nn\\
{x'}^- & = & -\frac{1}{p_-}\dot x^A {x'}^A + O(1/R^2)\ .
\ee

The conformal gauge constraints
themselves are only consistent with the equations of motion if the worldsheet
metric acquires curvature corrections (ie.~$h$ departs from the flat metric $h = {\rm diag}(-1,1)$), 
which we express symbolically as $\tilde h^{ab}$ according to
\be
h = \left(\begin{array}{cc}
	-1 + \tilde h^{\tau\tau}/R^2 & \tilde h^{\tau\sigma}/R^2 \\
	\tilde h^{\tau\sigma}/R^2 & 1 + \tilde h^{\sigma\sigma}/R^2
	\end{array} \right)\ .
\label{wsmetric}
\ee
The requirement that $\det h = -1$ implies $\tilde h^{\tau\tau} = \tilde h^{\sigma\sigma}$ and,
for future reference, the correction terms $\tilde h^{ab}$ are given explicitly to the order of interest
by
\be
\tilde h^{\tau\tau} &=& \frac{1}{2}(z^2-y^2) - \frac{1}{2p_-^2}\left[(\dot x^A)^2 + ({x'}^A)^2\right]\ ,
\nn\\
\tilde h^{\tau\sigma} &=& \frac{1}{p_-^2}\dot x^A {x'}^A\ .
\label{metriccorrections}
\ee
Finally, we note that the canonical momenta associated with the 
physical worldsheet excitations, defined by the variation $p_A = \delta {\cal L} / \delta x^A$, 
also acquire $O(1/R^2)$ corrections: consistent quantization requires that these 
corrections be taken into account.  Expressed in terms of canonical 
variables, the final bosonic Hamiltonian takes the form
\be
H_{\rm LC} &=& \frac{p_-}{2R^2}(x^A)^2 
	+ \frac{1}{2p_-R^2}\left((p_A)^2+({x'}^A)^2\right)
\nn\\
&&	+\frac{1}{R^4}\biggl\{
	\frac{1}{4p_-}\left[z^2(p_y^2+{y'}^2 +2{z'}^2) - y^2(p_z^2+{z'}^2+2{y'}^2)\right]
	+\frac{p_-}{8}\bigl[(x^A)^2\bigr]^2
\nn\\
&&\kern-30pt
	-\frac{1}{8p_-^3}\Bigl\{
	\bigl[(p_A)^2\bigr]^2+2(p_A)^2({x'}^A)^2 + \bigl[(x^A)^2\bigr]^2\Bigr\}
	+\frac{1}{2p_-^3}({x'}^A p_A)^2 \biggr\} +O(1/R^6)\ ,
\label{HLC}
\ee
where the pp-wave Hamiltonian emerges as expected at leading order.
The lightcone momentum $p_-$ is identified (via the AdS/CFT dictionary)
with the modified 't~Hooft parameter $\lambda'$ according to 
\be
p_- = 1/\sqrt{\lambda'} = J/\sqrt{\lambda}\ .
\ee

From the point of view of the semiclassical analysis, we are
working to two-loop order in quantum corrections.  
Since the quadratic theory
can be quantized exactly, however, we can study the quartic interaction Hamiltonian using 
standard first-order perturbation theory.  
A detailed analysis of the resulting spectrum of this perturbation can be 
found in \cite{McLoughlin:2004dh,Callan:2003xr,Callan:2004uv,Callan:2004ev}.  In the course of
those studies it was noticed that, analogous to the gauge theory closed
sectors studied in \cite{Beisert:2003tq,Beisert:2003jj,Beisert:2003jb,Beisert:2003ys}, certain
sectors emerged from the string analysis that decouple from the remainder 
of the theory to all orders in $\lambda'$.  One sector, which maps to the 
$\Sl(2)$ sector of the gauge theory, is diagonalized by 
bosonic string states excited in the $AdS_5$ subspace and forming symmetric-traceless
irreps in spacetime indices.  The corresponding sector of symmetric-traceless
$S^5$ string bosons maps to the closed $\su(2)$ sector in the gauge theory.
The block-diagonalization of these sectors in the string Hamiltonian will be
an important tool in the present analysis:  just as 
all higher hidden local charges in the gauge theory are simultaneously diagonalized
by a single Bethe ansatz, all of the higher hidden charges descending from the
string theory should be block-diagonalized by these particular string states as well.

\section{Lax representation}
The goal is to determine whether a ladder of higher local charges can be 
computed and quantized (albeit perturbatively), analogous to the existing treatment
of the near-pp-wave Hamiltonian given in eqn.~(\ref{HLC}) above.  
To quartic order in the semiclassical expansion defined by eqn.~(\ref{rescalePre}), the difference
between the string sigma model in eqns.~(\ref{action2a},\ref{action2b}) and that
of the $O(4,2)\times O(6)$ sigma model, defined by
\be
{\cal L}_{O(4,2)} & = &  -\frac{1}{2} \eta_{PQ}\del_a Z_P \del^a Z_Q  
	+ \frac{\tilde\varphi}{2} \left(\eta_{PQ}Z_P Z_Q + 1\right)\ , 
\nn\\
{\cal L}_{O(6)} & = &  -\frac{1}{2} \del_a Y_M \del^a Y_M  
	+ \frac{\varphi}{2} \left(Y_M Y_M - 1\right) \ ,
\label{O6action}
\ee
will essentially amount to an interaction perturbation due to curvature 
corrections to the worldsheet metric.  We therefore find it useful to rely on 
a known Lax representation of the $O(4,2)\times O(6)$ sigma model; this 
representation will define an unperturbed theory, and we will add perturbations
by hand to recover the full interaction Hamiltonian in eqn.~(\ref{HLC}).
(For a general introduction
to the Lax methodology in integrable systems, the reader is referred to \cite{Faddeev:ph}.)
Since worldsheet curvature corrections only appear at $O(1/R^2)$,
the reduction to the $O(4,2)\times O(6)$ sigma model at leading order in the 
expansion will be automatic.

For simplicity, we start from the four-dimensional Lax representation given 
for the $O(6)$ sigma model in \cite{Arutyunov:2003za} (see also \cite{Pohlmeyer:1975nb} 
for details), and work only to leading order in the semiclassical expansion.
The complexified coordinates 
\be
{\cal Y}_1 = Y_1 + i\,Y_2 \qquad {\cal Y}_2 = Y_3 + i\,Y_4
	\qquad {\cal Y}_3 = Y_5 + i\,Y_6\ 
\label{complexY}
\ee
are used to form a unitary matrix $S_{S^5}$
\be
S_{S^5} = \left(
	\begin{array}{cccc}
	0 & {\cal Y}_1 & -{\cal Y}_2 & \bar {\cal Y}_3 \\
	-{\cal Y}_1 & 0 &{\cal Y}_3 & \bar{\cal Y}_2 \\
	{\cal Y}_2 & - {\cal Y}_3 & 0 & \bar{\cal Y}_1 \\
	-\bar{\cal Y}_3 & -\bar {\cal Y}_2 & -\bar{\cal Y}_1 & 0 
	\end{array}
\right)\ ,
\label{smat}
\ee
in terms of which one may form the following $SU(4)$-valued currents:
\be
A_a = S_{S^5}\del_a {S_{S^5}}^\dag\ .
\ee
The equations of motion of the $O(6)$ sigma model 
\be
\del_a \del^a Y_M + \varphi Y_M = 0
\label{geneom}
\ee
are then encoded by
the auxiliary system of linear equations 
\be
(\del_\sigma - U)X = (\del_\tau - V)X = 0\ ,
\label{laxpair1}
\ee
where the Lax pair $U$ and $V$ are defined by
\be
U = \frac{1}{1+\gamma}A_- - \frac{1}{1-\gamma}A_+\ , \qquad 
\label{laxmat1}
V = -\frac{1}{1+\gamma}A_- - \frac{1}{1-\gamma}A_+\ .
\label{laxmat2}
\ee
The constant $\gamma$ is a free spectral parameter,
and $A_{\pm}$ are defined by $A_{\pm} \equiv \frac{1}{2}(A_\tau\pm A_\sigma)$.  
Note that on the $SO(4)$ subspace spanned by $y_{k'}$, eqn.~(\ref{geneom})
reduces to the pp-wave equations of motion on $S^5$:
\be
\ddot y_{k'} - {y''}_{k'} + p_-^2 y_{k'} = 0\ .
\label{ppeom}
\ee

The utility of the Lax representation arises from the fact that $U$ and $V$ may be 
considered as local connection coefficients, and a consistency equation for the
auxiliary linear problem can be reinterpreted as a flatness condition for the $(U,V)$-connection:
\be
\del_\tau U - \del_\sigma V + \left[U,V\right] = 0\ .
\label{ZCC1}
\ee
Parallel transport along this flat connection is defined by the path-ordered exponent
\be
\Omega_C(\gamma) = {\cal P}\exp \int_{\cal C} (U\,d\sigma + V\,d\tau) \ ,
\ee
where ${\cal C}$ is some contour in ${\bf R}^2$.
Restricting to transport along the contour defined by $\tau=\tau_0$ and $0\leq\sigma \leq 2\pi$
yields a monodromy matrix:
\be
T(2\pi,\gamma) = {\cal P}\exp \int_0^{2\pi} d\sigma\, U\ .
\label{monodromy}
\ee
The flatness condition in eqn.~(\ref{ZCC1}) admits an infinite number
of conservation laws, which translates to the fact that the trace of the monodromy matrix yields 
an infinite tower of local, mutually commuting charges $\Qo_n^{S^5}$ when expanded 
in powers of the spectral index about the poles of $U$ 
($\gamma = \pm 1$, in this case):\footnote{In general, an expansion around some $\gamma$
that is finitely displaced from a singularity of $U$ 
will yield combinations of local and non-local quantities.  One is of course free
to redefine $\gamma$ such that the expansion about $\gamma=0$ in eqn.~(\ref{chargedef})
is local. }
\be
\tr T(2\pi,\gamma) = \sum_n \gamma^n \widehat Q_n^{S^5}\ .
\label{chargedef}
\ee
The first nonvanishing charge $\Qo_2^{S^5}$, for example, is the Hamiltonian of the 
theory (on the $S^5$ subspace).

Moving beyond leading order in the semiclassical expansion, 
the essential difference between the $O(6)$ sigma model defined in 
eqn.~(\ref{O6action}) and the string action given in eqn.~(\ref{action2b}) 
is, as noted above, that worldsheet indices are contracted in the 
latter case with a non-flat worldsheet metric.  Keeping the components of
$h^{ab}$ explicit, the lightcone Hamiltonian derived from the string sigma model in 
eqn.~(\ref{action2b}) appears at leading order as
\be
H_{\rm LC}^{S^5} = -\frac{1}{2p_-R^2}\left[
	h^{\tau\tau}( p_-^2 y^2 + {y'}^2 + \dot y^2 )
	+2 h^{\tau\sigma} \dot y \cdot y' \right] + O(1/R^4)\ ,
\label{LOham}
\ee
where $h^{\tau\tau} = -1 + \tilde h^{\tau\tau}/R^2$ 
and $h^{\tau\sigma} = \tilde h^{\tau\sigma}/R^2$.
The prescription will be to find a perturbation to the 
$(U,V)$-connection such that the Hamiltonian in eqn.~(\ref{LOham})
emerges in an appropriate limit from the charge $\Qo_2^{S_5}$ defined by eqn.~(\ref{chargedef}).  
Such a perturbation is achieved by transforming the $U$ matrix according to 
\be
U\to U = \frac{1}{1+\gamma}(1+u_-/R^2 )A_- - \frac{1}{1-\gamma}(1+ u_+/R^2)A_+\ ,
\label{Udef}
\ee
where $u_\pm$ are given by
\be
u_\pm \equiv -\frac{1}{2} \tilde h^{\tau\tau} \mp \frac{1}{3} \tilde h^{\tau\sigma}\ . 
\ee
These perturbations should be treated as constants, to 
be replaced in the end with dynamical variables by fixing conformal gauge according to 
eqn.~(\ref{CGC}).  
The remaining quartic perturbations to the pp-wave theory will be naturally encoded in the
semiclassical expansion of the underlying $O(6)$ (likewise, $O(4,2)$) sigma model.
The matrix $V$ can be transformed in a similar way:
\be
V\to V = -\frac{1}{1+\gamma}(1+v_-/R^2)A_- - \frac{1}{1-\gamma}(1+v_-/R^2 )A_+\ ,
\ee
where $v_\pm$ may be chosen such that the perturbed Lax pair satisfies the flatness condition
in eqn.~(\ref{ZCC1}).  
Given that the intent is simply to determine whether the higher local charges generated by the 
perturbed monodromy matrix are conserved when quantum fluctuations are included, fixing $V$ to
satisfy the flatness condition is not really necessary:  the complicated formulas for $v_\pm$ 
that do satisfy eqn.~(\ref{ZCC1}) will therefore not be needed.

The perturbation in eqn.~(\ref{Udef}) can be obtained by a slightly different method.  When the path-ordered 
exponent defining the monodromy matrix is expanded, it can be seen that all odd products of 
the Lax matrix $U$ will not contribute to the final expression.  By replacing all even products
of $U$ according to the rule
\be
U(\sigma_1) U(\sigma_2) 
	&\to& \frac{1}{(\gamma^2-1)^2}\Bigl[
	h^{\sigma\sigma} A_\sigma(\sigma_1)A_\sigma(\sigma_2)
	-\gamma^2 h^{\sigma\tau} A_\sigma(\sigma_1)A_\tau(\sigma_2)
\nn\\
&&	-\gamma^2 h^{\tau\sigma} A_\tau(\sigma_1)A_\sigma(\sigma_2)
	-\gamma^2 h^{\tau\tau} A_\tau(\sigma_1)A_\tau(\sigma_2)
	\Bigr]\ ,
\label{UUrule}
\ee
the Hamiltonian in eqn.~(\ref{LOham}) is again obtained at leading order in the expansion.
Computationally, this latter method seems to be much more efficient, and
we will use eqn.~(\ref{UUrule}) in what follows.
At leading order in the $1/R$ expansion, the first nonvanishing integral of 
motion descending from the monodromy matrix is thereby found to be
\be
{Q}_2^{S^5} 
	= \frac{4\pi}{R^2}\int_0^{2\pi} d\sigma\,
	\left[
	h^{\tau\tau}( p_-^2 y^2 + {y'}^2 + \dot y^2 )
	+2 h^{\tau\sigma} \dot y \cdot y'\right] + O(1/R^4)\ ,
\label{Q20}
\ee
which, by construction, matches the desired structure in eqn.~(\ref{LOham}).

The same construction may be carried out for the $AdS_5$ system.
In fact, to make matters simple, we may borrow the Lax structure of the $O(6)$ 
model defined in eqns.~(\ref{smat}-\ref{laxmat2}), replacing the
$O(6)$ coordinates in eqn.~(\ref{complexY}) with the following Euclideanized 
$O(4,2)$ complex embedding coordinates:
\be
{\cal Z}_1 = Z_1 + i\,Z_2 \qquad {\cal Z}_2 = Z_3 + i\,Z_4
	\qquad {\cal Z}_3 = i\,Z_0 - Z_5\ .
\ee
In this case, however, the Lax matrix $S_{AdS_5}$ will obey $S_{AdS_5}^\dag S_{AdS_5} = -1$.
Otherwise, the analysis above applies to the $AdS_5$ sector by direct analogy: 
expanding the perturbed $O(4,2)$ monodromy matrix in the spectral parameter yields 
a set of charges labeled by $\widehat Q_n^{AdS_5}$.  The local charges
for the entire theory are then given by
\be
\widehat Q_n \equiv \widehat Q_n^{S^5} - \widehat Q_n^{AdS_5}\ .
\ee
The corresponding currents will be labeled by ${\cal Q}_n$.

It turns out that the expansion in the spectral parameter
$\gamma$ is arranged such that the path-ordered exponent defining
the monodromy matrix can be computed explicitly to a given order in
$\gamma$ by evaluating only a finite number of worldsheet integrals.  
The procedure for extracting local, canonically quantized currents is then
completely analogous to that followed in computing the
lightcone Hamiltonian described above.  All gauge fixing is done after the currents 
are evaluated, all occurrences of $x^-$ are replaced 
with dynamical variables by solving the conformal gauge constraints, and worldsheet 
metric corrections $\tilde h^{ab}$ are evaluated according to eqns.~(\ref{metriccorrections}) 
above.  We note, however, that previous studies 
involving the matching of integrable structures between gauge and string
theory have found it necessary to invoke certain redefinitions of $\gamma$ to obtain agreement
\cite{Beisert:2004hm,Kazakov:2004qf}.  
It would be straightforward to allow for rather general redefinitions of the spectral parameter
in the present calculation.
When we turn to computing spectra and comparing with gauge theory, however,
such redefinitions can lead to unwanted ambiguity.
We will therefore be primarily interested in finding ratios of 
eigenvalue coefficients for which arbitrary redefinitions of $\gamma$ are 
irrelevant, and for simplicity we will simply retain the original definition of $\gamma$
given by eqn.~(\ref{laxmat2}) above.

As previously noted, the first current ${\cal Q}_1$ defined by eqn.~(\ref{chargedef}) vanishes.  
In fact, all ${\cal Q}_n$ vanish for odd values of $n$, 
and this property of the integrable structure is mirrored 
on the gauge theory side.  The first nonvanishing current emerging from the monodromy matrix
is given by
\be
{\cal Q}_2 & = & \frac{4\pi }{R^2} \left( (\dot x^A)^2
    + ({x'}^A)^2 + p_-^2(x^A)^2 \right) 
\nonumber \\
& &     +\frac{\pi }{R^4}\biggl\{
     {2z^2} \left[{y'}^2
    + 2{z'}^2- \dot y^2 \right]
    - {2y^2} \left[{z'}^2 + 2{y'}^2
    - \dot z^2 \right]
	-\frac{4}{p_-^2}(\dot x^A{x'}^A)^2
\nonumber \\
& & 	+\frac{1}{p_-^2}\left[3(\dot x^A)^2-({x'}^A)^2\right]
	\left[(\dot x^A)^2+({x'}^A)^2\right]
	+{p_-^2}\left[(x^A)^2\right]^2 \biggr\} + O(1/R^6)\ .
\label{Q2pre}
\ee
The leading-order term is the quadratic pp-wave Hamiltonian, as expected, and the perturbation
is strictly quartic in field fluctuations.  All occurrences of $x^-$ and all curvature corrections 
to the worldsheet metric $\tilde h^{ab}$ have been replaced with physical variables as described above.
The final step is to express eqn.~(\ref{Q2pre}) in
terms of canonically conjugate variables determined by directly varying the Lagrangian in 
eqn.~(\ref{action1}).  We obtain
\be
\label{Hppwave}
{\cal Q}_2 & = & 
	\frac{4\pi }{R^2}\left({p_-^2}(x^A)^2 + (p_A)^2 + ({x'}^A)^2\right)
\nn\\
&&	+\frac{\pi }{R^4}\biggl\{
	{2}\left[ -y^2\left( p_z^2 + {z'}^2 + 2{y'}^2\right)
	+ z^2\left( p_{y}^2 + {y'}^2 + 2{z'}^2 \right)\right]
	+ {p_-^2}\left[ (x^A)^2 \right]^2
\nn\\
& &\kern-30pt
 	- \frac{1}{p_-^2}\Bigl\{  \left[ (p_A)^2\right]^2 + 2(p_A)^2({x'}^A)^2 
	+ \left[ ({x'}^A)^2\right]^2 \Bigr\}
	 + \frac{4}{p_-^2}\left({x'}^A p_A\right)^2
	\biggr\} + O(1/R^6)\ .
\label{hamQ2}
\ee
Comparing this with eqn.~(\ref{HLC}) above, we see that, to the order of interest,
\be
{\cal Q}_2 = 8\pi\, p_- H_{\rm LC}\ .
\ee
As expected, the perturbed monodromy matrix precisely reproduces the structure of 
the lightcone Hamiltonian to quartic order in the semiclassical expansion.  
(Note that ${\cal Q}_2$ is only expected to be identified with the lightcone Hamiltonian
up to an overall constant.)

Computationally, the expansion of the monodromy matrix becomes increasingly time consuming
at higher orders in the spectral index.  The situation can be mitigated to some extent 
by projecting the theory onto $AdS_5$ or $S^5$ excitations, eliminating all interaction
terms from the quartic perturbation that mix fluctuations from both subspaces.  
We will eventually want to compute eigenvalue spectra in the block-diagonal subsectors
discussed above (which require such a projection), so this maneuver will not affect the 
outcome.  

The next nonvanishing $S^5$ current in the series is given by
\be
{\cal Q}_4^{S^5} & = & \frac{8\pi}{3R^2}\left(3 - \pi^2 p_-^2 \right)
	\left(p_-^2 y^2 + p_y^2 + {y'}^2 \right)
\nn\\
&&	+\frac{2\pi}{3p_-^2 R^4}\biggl\{
	-3 
	( p_y^2 - 2p_y\cdot y' + {y'}^2 )
	( p_y^2 + 2p_y\cdot y' + {y'}^2 )
\nn\\
&&	-  \pi^2 p_-^2 \Bigl[ 4 (p_y\cdot y')^2 + (p_y^2 + {y'}^2)^2 \Bigr]
	- 12 p_-^2 {y'}^2 y^2
\nn\\
&&	- p_-^4 y^2 \Bigl[
		4 \pi^2 p_y^2 - 3  y^2\Bigr]
	-3 \pi^2 p_-^6 (y^2)^2 \biggr\}+ O(1/R^6)\ .
\label{hamQ4}
\ee
Although the quadratic interaction of ${\cal Q}_4^{S^5}$ is proportional to the pp-wave Hamiltonian on the
$S^5$, the structure of the perturbing quartic interaction differs from 
that obtained for ${\cal Q}_2$.  
The corresponding $AdS_5$ current takes the form
\be
{\cal Q}_4^{AdS_5} & = & \frac{8\pi}{3R^2}\left(3 - \pi^2 p_-^2 \right)
	\left(p_-^2 z^2 + p_z^2 + {z'}^2 \right)
\nn\\
&&
	+\frac{2\pi}{3p_-^2 R^4}\biggl\{ 
 	-3 
	( p_z^2 - 2p_z\cdot z' + {z'}^2 )
	( p_z^2 + 2p_z\cdot z' + {z'}^2 )
\nn\\
&&	-  \pi^2 p_-^2 \Bigl[ 4 (p_z\cdot z')^2 + (p_z^2 + {z'}^2)^2 \Bigr]
	+ 12 p_-^2 {z'}^2 z^2
\nn\\
&&	+ p_-^4 z^2 \Bigl[
		-4 \pi^2 {z'}^2 + 3  z^2\Bigr]
	+ \pi^2 p_-^6 (z^2)^2 \biggr\}+ O(1/R^6)\ ,
\label{hamQ4Z}
\ee
where the quadratic sector is again proportional to the pp-wave Hamiltonian, 
projected in this case onto the $AdS_5$ subspace.
Continuing on to sixth-order in the spectral index, 
we find the $S^5$ current
\be
{\cal Q}_6^{S^5} & = & \frac{1}{15R^2}\biggl\{
	4 \pi 
	\Bigl[
	45 -40 \pi^2 p_-^2 + 2\pi^4 p_-^4 \Bigr]
	\left(p_-^2 y^2 + p_y^2 + {y'}^2 \right) \biggr\}
\nn\\
&& 	+ \frac{\pi}{15p_-^2 R^4}\biggl\{
	 -45
	(p_y^2-2p_y\cdot y' +{y'}^2)( p_y^2+2 p_ y\cdot y' +{y'}^2)
\nn\\
&&	-20 p_-^2 \Bigl[
	2\pi^2 \left(4 (p_y\cdot y')^2 + (p_y^2+{y'}^2)^2\right)
	+9  {y'}^2 {y}^2 \Bigr]
\nn\\
&&	+ p_-^4\Bigl[
	2 \pi^4 \Bigl( 4 (p_y\cdot y')^2+3(p_y^2 +{y'}^2)^2\Bigr)
	-160 \pi^2  p_y^2 y^2 + 45 (y^2)^2 \Bigr]
\nn\\
&&\kern-10pt
	+8\pi^2 p_-^6 y^2\Bigl[
	(2\pi^2 p_y^2 + \pi^2 {y'}^2) - 15 y^2 \Bigr]
	+10 \pi^4 p_-^8 (y^2)^2
	\biggr\}+ O(1/R^6)\ .
\label{hamQ6}
\ee
The quadratic piece of ${\cal Q}_6^{S^5}$ is again identical in structure to the pp-wave Hamiltonian.
The analogous current in the $AdS_5$ subspace is arranged in a similar fashion:
\be
{\cal Q}_6^{AdS_5} & = & \frac{1}{15R^2}\biggl\{
	4 \pi \Bigl[
	45 -40 \pi^2 p_-^2 + 2\pi^4 p_-^4 \Bigr]
	\left(p_-^2 z^2 + p_z^2 + {z'}^2 \right) \biggr\}
\nn\\
&& 	+ \frac{\pi}{15p_-^2 R^4}\biggl\{
	 -45
	(p_z^2-2p_z\cdot z' +{z'}^2)( p_z^2+2 p_ z\cdot z' +{z'}^2)
\nn\\
&&	-20 p_-^2 \Bigl[
	2\pi^2  \left(4 (p_z\cdot z')^2 + (p_z^2+{z'}^2)^2\right)
	-9 {z'}^2 {z}^2 \Bigr]
\nn\\
&&	+ p_-^4\Bigl[
	2\pi^4 \Bigl( 4 (p_z\cdot z')^2+3(p_z^2 +{z'}^2)^2\Bigr)
	-160 \pi^2  {z'}^2 z^2 + 45 (z^2)^2 \Bigr]
\nn\\
&&\kern-10pt
	+8\pi^2 p_-^6 y^2
	(\pi^2 {z'}^2 + 5 z^2 )
	-6\pi^4 p_-^8 (z^2)^2
	\biggr\}+ O(1/R^6)\ .
\label{hamQ6Z}
\ee
While we will not present explicit formulas for the resulting currents, it is easy to carry this
out to eighth order in $\gamma$.

Taken separately, each current can be viewed as a free pp-wave Hamiltonian
plus a quartic interaction.  This is particularly useful, as it allows us to quantize each 
charge exactly at leading order and express the perturbation in terms of 
free pp-wave oscillators.  More explicitly, we quantize the quadratic sectors of these currents
by expanding the fluctuation fields in their usual Fourier components:
\be
x^A(\sigma,\tau) &=& \sum_{n=-\infty}^\infty x_n^A(\tau)e^{-ik_n\sigma}\ , \nn\\
x^A_n(\tau) &=& \frac{i}{\sqrt{2\omega_n}}\left( a_n^A e^{-i\omega_n\tau}
					+{a_n^{A\dag}} e^{i\omega_n\tau} \right)\ .
\ee
The quadratic (pp-wave) equations of motion
\be
\ddot x^{A} - {x''}^{A} + p_-^2 x^{A} = 0
\ee
are satisfied by setting $k_n = n$ (integer), and $\omega_n = \sqrt{p_-^2 + k_n^2}$, where 
the operators $a^A_n$ and ${a_n^{A\dag}}$ obey the commutation relation
$\bigl[ a_m^A,{a_n^{B\dag}}\bigr] = \delta_{mn}\delta^{AB}$.

In accordance with integrability, we expect that the local charges in 
eqns.~(\ref{hamQ2}--\ref{hamQ6Z}) 
should all be mutually commuting.  
Expressed in terms of quantum raising and lowering operators, we can check 
the commutators of the hidden local charges directly.  
To avoid mixing issues, We will need to select out closed subsectors of each charge which completely
decouple from the remaining terms in the theory.  We have already noted that the 
Hamiltonian $\Qo_2$ is known to be closed under $AdS_5$ and $S^5$ 
string states forming symmetric-traceless irreps in their spacetime indices.  
The equivalent gauge theory statement is that the dilatation generator is closed in
certain $\Sl(2)$ and $\su(2)$ projections.  Since the complete tower of corresponding charges in the 
gauge theory (including the dilatation generator) can be diagonalized by a single set of 
$\Sl(2)$ or $\su(2)$ Bethe equations, it is a reasonable guess that the full tower
of local string charges decouples under corresponding projections.  (A similar conjecture
is made, for example, in \cite{Arutyunov:2003rg,Kazakov:2004qf}.)  Following the 
treatment in \cite{McLoughlin:2004dh}, we therefore define the following $AdS_5$ oscillators
\be
a_n = \frac{1}{\sqrt{2}}\left(a_n^j + i a_n^k\right) \qquad 
\bar a_n = \frac{1}{\sqrt{2}}\left(a_n^j - i a_n^k\right) \qquad (j\neq k)\ ,
\label{oscads}
\ee
which satisfy the standard relations
\be
\left[ a_n, a_m^\dag \right] = \left[ \bar a_n, \bar a_m^\dag \right] = \delta_{nm} 
\qquad \left[ a_n, \bar a_m^\dag \right] = \left[ \bar a_n, a_m^\dag \right] = 0\ .
\ee
When restricted to these oscillators, the symmetric-traceless projection in the 
$AdS_5$ subspace is achieved by setting all $\bar a_n,\ \bar a_n^\dag$ to zero
(see \cite{McLoughlin:2004dh} for details).  
A corresponding definition on the $S^5$ takes the form
\be
a_n = \frac{1}{\sqrt{2}}\left(a_n^{j'} + i a_n^{k'}\right) \qquad 
\bar a_n = \frac{1}{\sqrt{2}}\left(a_n^{j'} - i a_n^{k'}\right) \qquad (j'\neq k')\ ,
\label{oscs5}
\ee
where the symmetric-traceless projection is again invoked by setting 
$\bar a_n,\ \bar a_n^\dag$ to zero.  In other words, we can test the commutativity
of the local charges in the $AdS_5$ and $S^5$ symmetric-traceless projections
by rewriting their oscillator expansions according to eqns.~(\ref{oscads},\ref{oscs5}) 
and setting all $\bar a_n,\ \bar a_n^\dag$ to zero.

Since the currents are expanded to $O(1/R^4)$, we only require that the commutators vanish 
to $O(1/R^6)$.  This simplifies the problem somewhat, since we only need to
compute commutators involving at most six oscillators.  On the subspace of symmetric-traceless 
$AdS_5$ string states, we obtain
\be
\left[ \widehat Q_n^{AdS_5},\widehat Q_m^{AdS_5} \right] 
= O(1/R^6) \qquad n,m \in 2,\ldots,8\ .
\ee
The corresponding projection on the $S^5$ yields
\be
\left[ \widehat Q_n^{S^5},\widehat Q_m^{S^5} \right] 
= O(1/R^6) \qquad n,m \in 2,\ldots,8\ .
\ee
We therefore find evidence for the existence of a tower of mutually 
commuting charges (within these particular closed sectors) that are 
conserved perturbatively by the quantized theory.

\section{Spectral comparison with gauge theory}
Given the freedom involved in redefinitions of the
spectral parameter, it may seem that any spectral agreement between the string charges computed above 
and corresponding quantities in the gauge theory would be rather arbitrary.  
We therefore seek a comparison of integrable structures on both sides of the
duality that avoids this ambiguity.  It turns out that such a test is indeed possible 
in the symmetric-traceless sector of $S^5$ excitations, which
will map in the gauge theory to the closed $\su(2)$ sector.  
We will further restrict ourselves to computing spectra associated with  
the following two-impurity string states:
\be
 {a_q^{j'\dag}}{a_{-q}^{k'\dag}} \ket{J}\ .
\nn
\ee
The analysis for three or higher-impurity states would require an accounting of interactions
between $AdS_5$ and $S^5$ string excitations;  as noted above, however, this dramatically 
complicates the computational analysis.  (We intend to return to the question
of higher-impurity string integrability in a future study.)
The ground state $\ket{J}$ is understood to carry $J$ units of angular momentum on the $S^5$,
and the two-impurity $SO(4)$ subspace above comprises a $16\times 16$-dimensional sub-block of the Hamiltonian. 
In addition, the mode indices (labeled here by $q$) of physical string states must sum to zero to
satisfy the usual level-matching condition
(the Virasoro constraint is understood to be satisfied by the leading-order solution
to the equations of motion; any higher-order information contained in the $T_{01}$ component
of eqn.~(\ref{CGC}) is irrelevant).

To simplify the analysis, and for comparison
with \cite{McLoughlin:2004dh,Callan:2003xr,Callan:2004uv,Callan:2004ev}, 
we will also rescale each of the charges computed above by a factor of $R^2$: 
\be
\widehat Q_n \to R^2 \widehat Q_n\ .
\ee
The two-impurity matrix elements of the charge $\widehat Q_2^{S^5}$ are then given by 
(note that the index notation is chosen here to align with 
\cite{McLoughlin:2004dh,Callan:2003xr,Callan:2004uv,Callan:2004ev}):
\be
\braket{J|a_q^{a'}a_{-q}^{b'}(\widehat Q_2^{S^5})a_{-q}^{c'\dag}a_{q}^{d'\dag} |J} & = & 
	{16\pi \omega_q} \delta^{a'd'}\delta^{b'c'}
\nn\\
&&\kern-80pt
	-\frac{8\pi q^2}{J \sqrt{\lambda'} \omega_q^2}
	\left[(3+2q^2\lambda)\delta^{a'd'}\delta^{b'c'} 
		-\delta^{a'c'}\delta^{b'd'} 
		+ \delta^{a'b'}\delta^{c'd'} \right]+ O(1/J^2)\ ,
\label{matq2}
\ee
where the upper spacetime indices $a',b',c',\ldots$ label the transverse $SO(4)$ 
directions in the $S^5$ subspace.
The radius $R$ has been replaced with the angular momentum $J$, and 
$p_-$ has been replaced with $1/\sqrt{\lambda'}$ via 
\be
J/p_- = R^2 = \sqrt{\lambda}\ .
\ee
As expected, contributions to the pp-wave limit of eqn.~(\ref{matq2}) all lie on the diagonal. 
Up to an overall factor, one may further check that the correction terms at $O(1/J)$ agree 
with those computed in \cite{Callan:2003xr,Callan:2004uv}, projected onto the $S^5$ subspace.  
The next higher charges in the series yield matrix elements given by
\be
\braket{J|a_q^{a'}a_{-q}^{b'}(\widehat Q_4^{S^5})a_{-q}^{c'\dag}a_{q}^{d'\dag} |J} & = & 
	\frac{32\pi\omega_q}{3\lambda'}( 3\lambda' - \pi^2  )\delta^{a'd'}\delta^{b'c'}
\nn\\
&&\kern-20pt
	-\frac{16\pi}{3{\lambda'}^{5/2}\omega_q^2 J}\biggl\{
	\Bigl[
	\pi^2(2+q^2\lambda') + 3 q^2 {\lambda'}^2(3+2q^2{\lambda'})\Bigr]\delta^{a'd'}\delta^{b'c'}
\nn\\
&&\kern-20pt
	+\Bigl[
	\pi^2(2+q^2\lambda')-3q^2{\lambda'}^2\Bigr]\delta^{a'c'}\delta^{b'd'}
\nn\\
&&\kern-20pt	
	+ q^2\lambda'(3\lambda' - \pi^2 ) \delta^{a'b'}\delta^{c'd'} 
	\biggr\}
	+ O(1/J^2)\ ,
\ee
\be
\braket{J|a_q^{a'}a_{-q}^{b'}(\widehat Q_6^{S^5})a_{-q}^{c'\dag}a_{q}^{d'\dag} |J} & = & 
	\frac{16\pi}{15\omega_q{\lambda'}^2}(2\pi^4 - 40\pi^2{\lambda'} + 45{\lambda'}^2 )
	\omega_q^2 \delta^{a'd'}\delta^{b'c'}
\nn\\
&&\kern-80pt
	+\frac{8\pi}{15{\lambda'}^{7/2}\omega_q^2}\biggl\{
	\Bigl[
	2  \pi^4 (4+q^2\lambda'(5+q^2\lambda'))
	-40\pi^2\lambda'(2+q^2\lambda')
\nn\\
&&\kern-80pt	
	-45 q^2{\lambda'}^3(3+2q^2\lambda') 
	\Bigr]\delta^{a'd'}\delta^{b'c'}
	+\Bigl[
	2 \pi^4(4+q^2\lambda')
	-40\pi^2\lambda'(2+q^2\lambda')
\nn\\
&&\kern-130pt
	+45q^2{\lambda'}^3  \Bigr]\delta^{a'c'}\delta^{b'd'}
	+q^2\lambda'\Bigl[
	\lambda'(40\pi^2  - 45\lambda' )-2\pi^4 \Bigr]\delta^{a'b'}\delta^{c'd'}
	\biggr\}+ O(1/J^2)\ .
\ee

We will again project onto 
symmetric-traceless irreps of $SO(4)\times SO(4)$, transforming
as $({\bf 1,1;3,3})$ in an $SU(2)^2\times SU(2)^2$ notation. 
Although it is not necessarily
guaranteed that the symmetric-traceless states will diagonalize the higher charges 
$\widehat Q_4$ and $\widehat Q_6$ at quartic order,
this can be checked directly at one-loop order in $\lambda'$ by computing the eigenvectors of the charges above
(the higher-loop version of this check is much more 
difficult because the above charges are no longer completely block diagonal under 
the $SO(4)$ projection, a fact that can be seen in the structure of ${\cal Q}_2$ above).
The $\widehat Q_2^{S^5}$ eigenvalue between symmetric-traceless $({\bf 1,1;3,3})$ 
$S^5$ states (denoted by $Q_2^{S^5}$) is then found to be
\be
Q_2^{S^5} = 16\pi  \left(\omega_q - \frac{q^2\sqrt{\lambda'}}{J}\right) + O(1/J^2)\ .
\label{Q2eig}
\ee
Up to an overall constant, this is just the two-impurity energy shift originally 
reported in \cite{Callan:2003xr,Callan:2004uv}.
The corresponding eigenvalues of the higher charges $\Qo_4$ and $\Qo_6$ can be
computed in an analogous fashion:
\be
Q_4^{S^5}  &= & \frac{32 \pi}{3}\biggl\{
	\frac{\omega_q}{\lambda'}
	\Bigl[ 3\lambda' - \pi^2  \Bigr]
\nn\\
&&\kern+00pt
	-\frac{\pi}{{\lambda'}^{5/2} \omega_q^2 J}\Bigl[
	\pi^2 (2+q^2\lambda') + 3q^2{\lambda'}^3\omega_q^2 \Bigr]
	\biggr\}+ O(1/J^2)\ ,
\label{Q4eig}
\nn\\
Q_6^{S^5} & = & \frac{16\pi}{15}\biggl\{
	\frac{\omega_q}{{\lambda'}^2}(2\pi^4 - 40 \pi^2\lambda' + 45{\lambda'}^2)
	-\frac{1}{\omega_q^2 {\lambda'}^{7/2} J}
	\Bigl[ 40 \pi^2\lambda'(2+q^2\lambda') 
\nn\\
&&
	+ 45 q^2{\lambda'}^4 \omega_q^2 - 2 \pi^4 (4+q^2\lambda' (3+q^2\lambda'))\Bigr]
	\biggr\} + O(1/J^2)\ .
\label{Q6eig}
\ee
Similar formulas can be extracted for the $AdS_5$ charges $Q_2^{AdS_5}$, 
$Q_4^{AdS_5}$ and $Q_6^{AdS_5}$, which are diagonalized by symmetric-traceless $({\bf 3,3;1,1})$
string states excited in the $AdS_5$ subspace.  Though we have not given explicit formulas,
it is also straightforward to obtain the corresponding eigenvalues for  $Q_8^{AdS_5}$ and $Q_8^{S^5}$.

By modifying the Inozemtsev spin chain of \cite{Serban:2004jf} to exhibit higher-loop BMN scaling, 
Beisert, Dippel and Staudacher were able to formulate a long-range
Bethe ansatz for the gauge theory in the closed $\su(2)$ sector \cite{Beisert:2004hm} 
(we will simply state their results here, referring the reader
to \cite{Beisert:2004hm} for further details).  In essence, the Bethe ansatz encodes 
the interactions of pseudoparticle excitations on a spin lattice and, in terms
of pseudoparticle momenta $p_k$, the ansatz given in \cite{Beisert:2004hm} diagonalizes the entire tower of local 
gauge theory $\su(2)$ charges.  The eigenvalues of these charges, which we label here 
as $D_n$, are given by
\be
D_n = \sum_{k=1}^M q_n (p_k)\ , \qquad 
q_n(p) = \frac{2\sin(\frac{p}{2}(n-1))}{n-1}
	\left( \frac{ \sqrt{1+8 g^2\sin^2(p/2)} -1}{2 g^2\sin^2(p/2)} \right)^{n-1}\ ,
\label{Ddef}
\ee
where $g^2 \equiv {\lambda}/{8\pi^2}$, and the index $k$ runs over the total number $M$ of 
pseudoparticle excitations (or $R$-charge impurities) on the spin lattice.
These eigenvalues can then be expanded perturbatively in
inverse powers of the gauge theory $R$-charge ${\cal R}$ by approximating the pseudoparticle
momenta $p_k$ by the expansion
\be
p_k = \sum_j \frac{f_j(n_k)}{{\cal R}^{j/2}}\ ,
\ee
where $f_j$ are functions of the integer mode numbers $n_k$, determined by solving the
Bethe equations explicitly to a given order in $1/{\cal R}$. 

In general, we wish to identify the local string charges with linear combinations of 
corresponding charges in the gauge theory.  From eqn.~(\ref{Ddef}), however, it is easy to see that
as one moves up the ladder of higher charges in the gauge theory, the eigenvalues $D_n$ of these
charges have leading contributions at higher and higher powers of $g^2/{\cal R}^2$
in the large-${\cal R}$, small-$\lambda$ double-scaling expansion.  This is puzzling because the
string eigenvalues computed above do not exhibit similar properties.
The difference in scaling behavior therefore motivates the following prescription 
for identifying the eigenvalues of the higher local charges on both sides of the correspondence:
\be
Q_{n} - N = C \left(\frac{n}{2} D_{n}\right)^{2/n}\ .
\label{QdefD}
\ee
$N$ here counts the number of string worldsheet impurities and $C$ is an arbitrary constant.  
Fractional powers of the gauge theory charges $D_n$ 
are well defined in terms of the double-scaling expansion, so that the right-hand side of 
eqn.~(\ref{QdefD}) is in fact just a linear combination of conserved quantities 
in the gauge theory. 

A potential subtlety arises when matching $Q_n$ and 
$D_n$ in this fashion for $n>2$ beyond one-loop order in $\lambda$.  
The problem is that, under the identification in eqn.~(\ref{QdefD}), information 
from string energy eigenvalues at $O(1/J^2)$ and higher
is required to completely characterize the higher-loop (in $\lambda$) coefficients
of the gauge theory charges $D_n$.  
The essential reason for this is that the string loop expansion is in powers 
of the modified 't~Hooft coupling which, in terms of the gauge theory $R$-charge ${\cal R}$, is
\be
\lambda' = \lambda/J^2 = \lambda/{\cal R}^2\ .
\label{RJid}
\ee
In other words, under eqn.~(\ref{QdefD}), it is
impossible to disentangle higher-order $1/J$ contributions to the string charges 
$Q_n$ from higher-order $\lambda$ corrections to $D_n$.
The prescription given in eqn.~(\ref{QdefD}) therefore holds only 
to one-loop order in $\lambda$, where knowing the $1/J$ corrections in the string
theory is sufficient.

Furthermore, since the local charges in the string and gauge theories are only 
identified up to an overall multiplicative constant, directly comparing the spectra 
of each theory is not especially rigorous.  A convenient quantity to work 
with, however, is the ratio of the $O(1/J)$ eigenvalue correction to the pp-wave coefficient:  
at first loop order in $\lambda$ this ratio eliminates all ambiguity 
associated with overall constants and $\gamma $ redefinitions, and 
thus provides a meaningful comparison with gauge theory. 
(The analogous quantity computed for charges in the $AdS_5$ subspace is not free from such
ambiguities.)
We therefore arrange the one-loop, two-impurity 
eigenvalues of local $S^5$ string theory charges according to
\be
Q_n^{S^5} = 2 + q^2 \lambda'\left(\Lambda_{n,0} + \frac{\Lambda_{n,1}}{J} \right) + O({\lambda'}^2) + O(1/J^2)\ ,
\ee
where the numbers $\Lambda_{n,0}$ and $\Lambda_{n,1}$ characterize eigenvalue coefficients in the
pp-wave limit and at $O(1/J)$, respectively, and $q$ is the mode number associated with the two-impurity
string states defined above.  
On the gauge theory side we make a similar arrangement:
\be
\left(\frac{n}{2} D_{n}\right)^{2/n} 
	=  \frac{q^2 \lambda}{{\cal R}^2}\left(\bar\Lambda_{n,0} + \frac{\bar\Lambda_{n,1}}{{\cal R}} \right) 
							+ O({\lambda}^2) + O(1/{\cal R}^4)\ ,
\ee
where the integer $q$ is a mode number associated with the momenta of pseudoparticle
excitations on the spin lattice (which, in turn, correspond to roots of the $\su(2)$ Bethe equations).
The $R$-charge ${\cal R}$ is understood to be identified with the string angular momentum $J$ via eqn.~(\ref{RJid}).

\begin{table}[ht!]
\begin{eqnarray}
\begin{array}{|c|cc|}
\hline
n & \Lambda_{n,1}/\Lambda_{n,0} & \bar\Lambda_{n,1}/\bar\Lambda_{n,0}  \\
\hline
2 & -2 & -2 \\
3 & 0 & 0 \\
4 & -2 & -2 \\
5 & 0 & 0 \\
6 & -2 & -2 \\
7 & 0 & 0 \\
8 & -2 & -2 \\
\hline
\end{array} \nonumber
\end{eqnarray}
\caption{Ratios of $O(1/J)$ (or $O(1/{\cal R})$) 
	corrections to pp-wave/BMN coefficients in string and gauge theory local charges.}
\label{tab1}
\end{table}

The quantities $\Lambda_{2,0}$ and $\Lambda_{2,1}$ 
for the string Hamiltonian $Q_2$ can be computed from the eigenvalue formula
in eqn.~(\ref{Q2eig}) (or, alternatively, 
retrieved from the two-impurity string results reported in \cite{Callan:2003xr,Callan:2004uv}).
We find the following ratio:
\be
\Lambda_{2,1}/\Lambda_{2,0} = -2\ .
\ee
As shown in \cite{Callan:2003xr,Callan:2004uv}, this agrees with the corresponding gauge theory
prediction at one-loop order in $\lambda$:
\be
\bar\Lambda_{2,1}/\bar\Lambda_{2,0} = -2\ .
\ee
The ratio of $O(1/J)$ eigenvalue corrections to pp-wave coefficients is in fact 
$-2$ for all of the nonvanishing string charges.
Under the matching prescription in eqn.~(\ref{QdefD}), this agrees with the gauge theory perfectly.  
(The odd charges vanish altogether on both sides of the correspondence.) 
We summarize the results of this comparison for the first eight charges in the series in Table~\ref{tab1}. 
It would be satisfying to test this agreement at higher loop-orders in $\lambda$.  The corresponding
computation at two-loop order, however, would require evaluating the local string theory charges
at $O(1/R^6)$ in the semiclassical expansion, where several subtleties of perturbation theory 
(and, for that matter, lightcone quantization) would need to be addressed.
This emphasizes the need to understand the quantum string theory at higher orders in the
expansion away from the pp-wave limit.

\section{Discussion}
We have provided evidence that an infinite tower of local,
mutually commuting bosonic charges of type IIB string theory on $AdS_5\times S^5$,
known to exist in the classical theory, can also be identified in the quantum theory.  
In addition, we have provided a prescription
for matching certain eigenvalues of these charges in a protected subsector
of the string theory to corresponding eigenvalues in the closed $\su(2)$ sector
of the gauge theory.  
The fact that the spectra of local string charges computed here can only be matched 
to corresponding quantities in the gauge theory via the matching prescription in
eqn.~(\ref{QdefD}), however, indicates that the monodromy matrix used 
to derive the local string charges is substantially different from that which would give rise
to the proposed quantum string Bethe ansatz of \cite{Arutyunov:2004vx} (or, since they are equivalent at
one-loop order, the corresponding $\su(2)$ Bethe ansatz in the gauge theory).  
In other words, we expect that there is a Lax representation for the string 
sigma model that gives rise to hidden local charges that can be compared directly 
with the gauge theory, without having to take fractional powers or linear combinations.  

There are a number of additional tests of integrability in the quantum 
string theory which, in the context of the present calculation,
should be relatively straightforward.  By computing the quartic interactions
among fluctuations in the $AdS_5$ and $S^5$ subspaces for each of the 
higher local charges studied here, it would be easy, for example, to find the
resulting spectra of three- or higher-impurity string states.  
Apart from the difficulty of actually computing the mixing interactions, 
this would provide a simple check on the methodology employed here.
A more difficult problem would be to address whether the integrable structure
of the string theory respects supersymmetry.  By formulating a supersymmetric
Lax representation that generates the complete interaction Hamiltonian computed
in \cite{Callan:2003xr,Callan:2004uv}, one might be able to show that each of the higher
local charges are individually supersymmetric, and a comparison with gauge theory
could be carried out in the closed $\su(1|1)$ sector studied in \cite{McLoughlin:2004dh,Callan:2004ev} 
(the corresponding sector of the string theory would be comprised of symmetrized
fermionic excitations in the $({\bf 3,1;3,1})$ or $({\bf 1,3;1,3})$ of
$SO(4)\times SO(4)$).  

Ultimately, the hope is that the arsenal of techniques associated with 
integrable systems will lead to an exact solution 
to the string formulation of large-$N_c$ Yang-Mills theory.  
Alternatively, a proof that both sides of the duality are diagonalized by 
identical Bethe equations should be obtainable.  At present, the major obstacle preventing such a
proof is the disagreement between gauge and string theory at three-loop order in the
't~Hooft coupling.   The fact that the integrable
systems of both theories seem to agree in certain limited cases, however, 
stands as strong evidence that they are likely to be equivalent.

\section*{Acknowledgments}
We would like to thank Curtis Callan for guidance and for reading the manuscript.
We also thank Sergey Frolov, Vladimir Kazakov, Tristan McLoughlin, Andrei Mikhailov, 
John Schwarz and 
Arkady Tseytlin for many helpful discussions, The Ohio State University 
and the Institute for Advanced Study for hospitality,
and the James A.~Cullen Memorial Fund for support.  This work was supported in part 
by US Department of Energy grant DE-FG03-92-ER40701.






\end{document}